\documentclass[%
 reprint,
 amsmath,amssymb,
 aps,
]{revtex4-2}

\usepackage{graphicx}
\usepackage[bb=boondox]{mathalfa}
\usepackage{dcolumn}
\usepackage{bm}
\usepackage{subcaption}
\usepackage{amsmath}
\usepackage{mathtools}
\usepackage{physics}
\usepackage{comment}
\usepackage[normalem]{ulem}
\usepackage{tikz-network}
\usepackage{ragged2e} 
\usepackage{comment}
\usepackage{todonotes}
\usetikzlibrary{decorations.pathmorphing}
\DeclareCaptionJustification{justified}{\justifying}

\usepackage{hyperref}
\hypersetup{
	colorlinks=true,
	linkcolor=blue,
	citecolor=blue,
	filecolor=magenta,
	urlcolor=cyan
}

\begin{document}

\title{Zero-point energy of tensor fluctuations on the MPS manifold}

\def\UCL{{London Centre for Nanotechnology, University College London,
Gordon St., London WC1H 0AH, United Kingdom}}

\author{Sebastian Leontica}
\email{sebastian.leontica.22@ucl.ac.uk}
\affiliation{\UCL}
\author{Andrew G. Green}
\affiliation{\UCL}

\date{\today}

\begin{abstract}
This work presents a method for studying low-energy physics in highly correlated magnetic systems using the matrix product state (MPS) manifold. We adapt the spin-wave approach, which has been very successful in modeling certain low-entanglement magnetic materials, to systems where the ground state is better represented by an MPS, such as the S = 1 Affleck-Kennedy-Lieb-Tasaki (AKLT) model. We argue that the quasi-local action of tensor fluctuations and the natural Kähler structure of the MPS manifold facilitate a description in terms of bosonic modes. We apply this approach to compute fluctuation corrections to the bilinear-biquadratic Heisenberg model, whose ground state we expect to be close to the exact bond dimension 2 AKLT state in a certain parameter range. Our results show significant improvements in energy approximations, highlighting both the qualitative and quantitative potential of this paradigm for studying complex entangled systems. This approach paves the way for new insights into the low-energy physics of correlated materials and the development of effective field theories beyond traditional semiclassical methods.
\end{abstract}

\maketitle

\textit{Introduction}---The theory of spin-waves, first introduced by Bloch \cite{Bloch1930}, is a very important tool for understanding the behaviour of low-temperature magnetic materials. At its core, it consists of assuming an ordered product state background whose excitations are described by propagating fluctuations of the spin direction. If the fluctuations are sufficiently weak, they can be modelled using bosonic modes via the Holstein-Primakoff transformation \cite{Holstein1940}. This has been extremely effective at explaining a range of phenomena in quantum systems with ferromagnetic and antiferromagnetic orderings \cite{Anderson1952, Kubo1952, Manousakis1991, Haldane1983, Auerbach1988}. 

In more recent years, tensor networks (and matrix product states more specifically for the 1D case) have been among the most widely used tools in condensed matter theory \cite{Orus2014,Verstraete2008, Cirac2009}, giving rise to both numerous theoretical results such as a classification of possible phases of matter in 1D \cite{Schuch2011} as well as unmatched numerical accuracy in tasks such as ground state approximations \cite{White1992, White1993, Schollwock2005, Schollwock2011}. The proliferation of these techniques stems from the ability to represent area-law entanglement patterns using only polynomially many parameters, which is enough to faithfully reproduce the ground state of gapped local Hamiltonians. Despite their success, understanding the low-energy physics of such systems is still an area of active research. The reason for this is that the dynamics following a global quench is expected to thermalize in most systems, leading to volume law entanglement and a breakdown of the matrix product state representation. A number of techniques have been proposed to tackle this issue and gain insight into low energy physics, such as tensor network based variational ansatze for understanding excitation spectra \cite{Haegeman2012, Haegeman2013(2), Vanderstraeten2019, Vanderstraeten2018} and the truncation of operator entanglement for computing transport properties \cite{Rakovszky2022}.

We propose here a method for understanding the low-energy physics based on an adaptation of the spin-wave treatment to magnetic systems whose ground states have more complex entanglement patterns than simple ferro- or antiferromagnetic ordering, such as the $S=1$ Affleck-Kennedy-Lieb-Tasaki (AKLT) model \cite{Affleck1987}. The low-energy physics is still governed by propagating waves treated as independent bosonic modes, but the local disturbance is not in the orientation of the spins, but rather a fluctuation of the tensors in the ground state MPS approximation.


Understanding the MPS double tangent space, as introduced and studied in previous works \cite{Haegeman2013, Haegeman2014}, was fundamental to this research. The Configuration Interaction with Single and Double excitations (CISD) ansatz for approximating ground states was discussed earlier in \cite{Wouters2013}, where the generalized eigenvalue equation in the redundant space of single and double tangent space vectors was explicitly constructed and solved.
Previous work\cite{Green2016} attempted to subsume these ideas into a field theoretical framework by constructing a path integral on the MPS manifold. Expanding about the saddles of this theory suggested a bosonic interpretation for the CISD equation and derived an effective 
Hamiltonian for the excitations. The normal part is identical to that used in the excitation ansatz \cite{Haegeman2012}, but additionally includes anomalous terms accounting for production of bosons in pairs. Care must be taken with this expansion, however. A naive treatment overcounts the number of bosonic states. We complete this picture by incorporating the insights of Ref.\cite{Leontica2024} and explicitly constructing a quasi-local bosonic Gram operator to recover the correct overlaps between the various tangent spaces, resolving previously observed overcounting of degrees of freedom. We note that this approach offers significant technical and philosophical advantages over brute-force optimization methods.

\textit{Matrix product states}---For a system with $N$ sites and open boundary conditions, we introduce the generic form of a matrix product state (MPS) defined via the local 3-tensors $\mathcal{A} = \{A^{[i]}\}_{i=1}^N$ with a physical index of dimension $d$ and two bond indices going up to the bond dimension $D$. The bond indices are contracted sequentially to produce the matrix product state

\begin{equation}
\begin{split}
    \ket{\Psi[\mathcal{A}]} &= \sum_{\{n_i\}} A^{[1]}_{n_1}A^{[2]}_{n_2} \dots A^{[N]}_{n_N} \ket{n_1 n_2 \dots n_N}, \\
    &= \begin{tikzpicture}[baseline={([yshift=1ex]current bounding box.center)}]
        \Vertex[x = 0.2, y = 0.5, size = 0.7, style = {color=white}]{O1}
        \Vertex[label = $A^{[1]}$,size = 0.7]{A} \Vertex[x=1,label = $A^{[2]}$,size = 0.7]{C} \Vertex[y=-1,style = {color=white}]{D} \Vertex[y=-1,x = 1,style = {color=white}]{B} \Vertex[x=2,style = {color=white}]{F}
        \Edge(A)(C)
        \Edge(A)(D)
        \Edge(C)(B)
        \Edge(C)(F)
    \end{tikzpicture} \ldots \begin{tikzpicture}[baseline={([yshift=1ex]current bounding box.center)}]
        \Vertex[x = -0.2, y = 0.5, size = 0.7, style = {color=white}]{O2}
        \Vertex[label = $A^{[N]}$,size = 0.7]{A} \Vertex[x=-1,style = {color=white}]{C} \Vertex[y=-1,style = {color=white}]{D}
        \Edge(A)(C)
        \Edge(A)(D)
    \end{tikzpicture}.
\end{split}
\end{equation}
Such constructions are known to be subject to gauge freedom, as it can easily be seen that any transformation of the type $A^{[i]}\to A^{[i]}X$, $A^{[i+1]}\to X^{-1}A^{[i+1]}$ will give rise to the same state on the physical legs. It is known that this issue can be resolved without losing expressivity by enforcing a more restrictive parameterization for the tensors at each site. Part of the gauge is fixed by enforcing left-canonical form, which can be obtained from a unitary $U$ in the following way
\begin{equation}
\label{eq:AfromU}
    \begin{tikzpicture}[baseline={([yshift=1ex]current bounding box.center)}]
        \Vertex[label = $A^{[i]}$,size = 0.7]{A} \Vertex[x = 1,style = {color=white}]{B} \Vertex[x=-1,style = {color=white}]{C} \Vertex[y=-1,style = {color=white}]{D}
        \Edge(A)(B)
        \Edge(A)(C)
        \Edge(A)(D)
    \end{tikzpicture} = 
    \begin{tikzpicture}[baseline={([yshift=1ex]current bounding box.center)}]
        \Vertex[label = $U^{[i]}$,shape = rectangle,size = 0.7]{A} \Vertex[x = 1,style = {color=white}]{B} \Vertex[x=-1,style = {color=white}]{C} \Vertex[y=-1,style = {color=white}]{D} \Vertex[y=0.9, label = 0, opacity =0,size=0.4]{F}
        \Edge[Direct](B)(A)
        \Edge[Direct](A)(C)
        \Edge[Direct](A)(D)
        \Edge[Direct](F)(A)
    \end{tikzpicture}.
\end{equation}

The remaining gauge freedom is fixed by restricting $U$ to the following parameterized form $D\times d$ unitary $U^{[i]}(x) = U\exp \left( x_{i} \bra{0}\otimes\Gamma^{1/2} -  \ket{0}\otimes\Gamma^{1/2} x^\dagger_{i}\right)$. This can be shown to be identical to the parameterization proposed in \cite{Wouters2013}, which covers the entire MPS manifold without gauge freedom. The above is a local parameterization around some reference unitary $U$, with the exponent given by the $\mathfrak{su}(D\times d)$ algebra element
\begin{equation}
    \begin{tikzpicture}[baseline={(current bounding box.center)}]
        \Vertex[style = {color=white}]{A}
        \Vertex[y=-1,style = {color=white}]{B}
        \Vertex[x=1,y=-1,RGB,color = {190, 221, 186},label = $x_i$]{C}
        \Vertex[label = $\Gamma^{-\frac{1}{2}}$,size = 0.7,x=2,y=-1,RGB,color={253,192,134}]{D}
        \Vertex[label = $0$,size = 0.4,x = 2,color=white]{E}
        \Vertex[x = 3,style = {color=white}]{F}
        \Vertex[x = 3,y=-1,style = {color=white}]{G}
        \Vertex[x=0.8,y=0.1,size = 0, label = $\delta \neq 0$,style = {color=white}]{H}
        \Edge[bend=45](A)(C)
        \Edge(B)(C)
        \Edge(C)(D)
        \Edge(E)(F)
        \Edge(D)(G)
    \end{tikzpicture} - \operatorname{h.c.},
\end{equation}
where the entries of the $D\times D\times (d-1)$ tensor $x_i$ are arbitrary complex numbers, and $\Gamma$ is the right environment. In the following, we will assume that the MPS is translationally invariant ($U$ has no site dependence) and its correlation length $\xi$ is finite.

This choice guarantees that all (sufficiently small) values of the parameters $x_i$ give rise to distinct physical states. We will use the Greek letters $\mu, \nu $, etc. to index the different entries of $x_i$, corresponding to the different ways in which the local tensor can fluctuate. The advantage of this choice lies in having the simple Gram matrix for the vectors in the first tangent space $\bra{\partial_\mu^{(i)}\Psi}\ket{\partial_\nu^{(j)}\Psi} = \delta_{\mu\nu}\delta_{ij}$ in the thermodynamic limit $N\to \infty$ \cite{Haegeman2013}. We will call the presence of a derivative on some site a local excitation, as it can be seen that the state $\ket{\partial_\mu^{(i)}\Psi}$ is also an MPS, where the tensor at site $i$ was replaced by a tangent tensor $B_\mu$ given by
\begin{equation}
\label{eq:derivative}
    \begin{tikzpicture}[baseline={([yshift=2ex]current bounding box.center)}]
        \Vertex[label = $B_\mu$,size = 0.7]{A} \Vertex[x = 1,style = {color=white}]{B} \Vertex[x=-1,style = {color=white}]{C} \Vertex[y=-1,style = {color=white}]{D}
        \Edge(A)(B)
        \Edge(A)(C)
        \Edge(A)(D)
    \end{tikzpicture} = 
    \begin{tikzpicture}[baseline={([yshift=1ex]current bounding box.center)}]
        \Vertex[label = $U$,shape = rectangle,size = 0.7]{A} \Vertex[label = $e_\mu$,x = 1,RGB,color = {190, 221, 186}]{B} \Vertex[x=-1,style = {color=white}]{C} \Vertex[y=-1,style = {color=white}]{D} 
        \Vertex[label = $\Gamma^{-\frac{1}{2}}$,size = 0.7,x = 2,RGB,color={253,192,134}]{G} \Vertex[x=3, style = {color=white}]{H}
        \Edge[Direct](B)(A)
        \Edge[Direct](A)(C)
        \Edge[Direct](A)(D)
        \Edge(G)(B)
        \Edge[bend = -90,Direct](B)(A)
        \Edge(H)(G)
    \end{tikzpicture},
\end{equation}
where $e_\mu$ is a unit tensor with a 1 in position $\mu$ and 0 elsewhere. Vectors spanning higher tangent spaces can be obtained by taking derivatives at different sites.

\textit{Bosonic mapping}---
In the following section we will argue that the local perturbations of the reference MPS can be understood {\it via} a set of  bosonic modes. These excitations are produced by taking derivatives of the MPS along different directions $\mu$ at different sites $i$ and can be associated with creation and annihilation operators $a_\mu^{(i)\dagger}$, $a_\mu^{(i)}$.

In the case of single excitations (or none) the orthonormalization of the single tangent space ensures that the overlaps are identical to those obtained from a bosonic Fock space. In the case of two or more excitations, this identification holds only when the separation is much larger than the correlation length. However, complications arise due to an overcounting of states with excitations separated by small distances. One can see this by noting that the basis formed by taking derivatives will have $(1+D^2(d-1))^N$ states, corresponding to placing either an $A$ tensor or one of the $B_\mu$ tensors at each site. If $D>1$, this will be larger than $d^N$, the dimension of the physical Hilbert space, indicating redundancy in the bosonic framework that must be addressed. It is a direct consequences of the delocalized nature of excitations of entangled reference states.

We give a general construction accounting for these effects for excitations above an MPS reference state. First, we consider
the (unnormalized) state $|\partial_\mu^{(i)}\partial_\nu^{(i+x)}\Psi \rangle $ where the separation is large compared to the correlation length of the reference state $x \gg \xi$. We will look at the overlap of this with another two excitation state $|\partial_\alpha^{(j)}\partial_\beta^{(j+y)}\Psi\rangle $. The left canonicalization of the MPS ensures that for a non-zero overlap we need to take the left-most derivatives in the same place, so that $i=j$. If we introduce the notation $T(A,B)$ for the transfer matrix
\begin{equation}
\label{eq:rightenv}
    T(A,B) = \begin{tikzpicture}[baseline={([yshift=-0.6ex]current bounding box.center)}]
        \Vertex[label = $A$,size = 0.7]{A} \Vertex[x=-1,style = {color=white}]{C} \Vertex[y=-1, label = $B^*$,size = 0.7]{D} \Vertex[x=-1,y=-1,style = {color=white}]{F} \Vertex[x=1,style = {color=white}]{G}
        \Vertex[x=1,y = -1,style = {color=white}]{H}
        \Edge(A)(D)
        \Edge(A)(C)
        \Edge(D)(F)
        \Edge(A)(G)
        \Edge(D)(H)
    \end{tikzpicture},
\end{equation}
and the transfer matrix of the reference state $T(A,A) = T$ we see that the overlap between the states with $y<x$ is
\begin{equation}
\begin{split}&
\langle \partial_\alpha^{(i)}\partial_\beta^{(i+y)}\Psi
|
\partial_\mu^{(i)}\partial_\nu^{(i+x)}\Psi
\rangle
\\
&= 
\left(I|
T(B_\mu, B_\alpha) T^{y-1}T(A, B_\beta)T^{x-y-1}T(B_\nu,A)
| \Gamma \right),
\end{split}
\end{equation}
where we have used the fact that $\lim_{n\to \infty} T^n = 
| \Gamma )(I|$ and follow the notation of Ref.\cite{Haegeman2013} for states in the doubled auxilliary space acted upon by the transfer matrix.
When $x \gg \xi$, one of the two powers of the transfer matrix in the above expression will be close to its limit at infinity, which is null since $(I| T(A,B_\mu) = 0$ by the definition of the tangent space tensors. The result of the overlap above is then guaranteed to be of order $\mathcal{O}(e^{-x/\xi})$. When the excitations on the right-most site are also directly on top of each other we have instead
\begin{equation}
\begin{split}&
\langle \partial_\alpha^{(i)}\partial_\beta^{(i+x)}\Psi|\partial_\mu^{(i)}\partial_\nu^{(i+x)}\Psi \rangle
\\
&= (I|T(B_\mu, B_\alpha) T^{x-1}T(B_\nu,B_\beta)|\Gamma) \\
&\approx (I|T(B_\mu, B_\alpha) |\Gamma) (I|T(B_\nu,B_\beta)|\Gamma)= \delta_{\mu\alpha}\delta_{\nu\beta},
\end{split}
\end{equation}
which is the same as we would expect from bosonic excitations, up to the small error which is exponentially suppressed in the distance. This argument  allows us to break overlaps into clusters whenever there is a separation much larger than $\xi$ between the excitations. When the excitations are located close together, the argument above fails and we need to account for the deviations from bosonic overlaps. 

In order to deal with the non-trivial overlaps of nearby excitations, we introduce the Gram operator $\mathcal{G}$, which we define such that
\begin{equation}
\label{eq:innerprod}
\begin{split}
\bra{\prod_{\mu,i}\partial_\mu^{(i)}\Psi} \ket{
\prod_{\nu,j}\partial_\nu^{(j)}\Psi}& \\
= [GS|\prod_{\mu,i}&a_\mu^{(i)} \mathcal{G} \prod_{\nu,j}a_\nu^{\dagger(j)}|GS],
\end{split}
\end{equation}
for all possible sets of positions and directions of the excitations (restricted to a single excitation per site). The notation $|\cdot]$ is used to distinguish states in the bosonic space from those in the physical Hilbert space, with a more detailed construction given in the Supplementary Material. We can expand the Gram operator as a normally ordered polynomial of the creation and annihilation operator, with the argument above guaranteeing that each monomial either has no gaps larger than $\sim \xi$ or is exponentially suppressed. 
After performing a change of coordinates such as to remove the overlap between the first and second tangent spaces (details in the Supplementary Material) 
the Gram operator can be expanded as

\begin{equation}
    \mathcal{G} = 1 +  \sum_{\substack{\mu\nu\alpha\beta,i\\x,y>0}}g_{\mu\nu\alpha\beta}^{(xy)}a_\mu^{(i)\dagger} a_\nu^{(i+x)\dagger} a_\alpha^{(i)} a_\beta^{(i+y)} + \ldots,
\end{equation}
where $g$ becomes exponentially small for large $x$ or $y$. 
The Gram operator also encodes constraints related to which excitations are allowed to be near each other. This is seen from the fact that $\mathcal{G}$ is not full rank, so there are two-excitation states of zero norm under $\mathcal{G}$. On the other hand, some two-excitation states obtain much larger norms than would be predicted by the bosonic analogy, giving rise to the overcounting that was discovered in an earlier work \cite{Leontica2024}.

\textit{Operators}---In this section we will discuss two approaches for obtaining the bosonic representation of some operator acting on the physical space. Such representations are similar to the Holstein-Primakoff transformation, with two important distinctions. The expansion of an operator at site $i$ in bosonic modes is only quasi-local, with some contribution from bosonic operators located at sites to the right of $i$. Secondly, for a faithful bosonic representation we need to cut out not only those states with too many excitations per site, but also all states of zero norm under the inner product defined in Eq.~\eqref{eq:innerprod}. A more rigorous introduction to the mapping is given in the Supplementary Material.

We will mainly be referring to the Hamiltonian $H$, but the same construction holds for any operator of interest. The first method is based on tensor network contractions and has previously been used by Haegeman et al. \cite{Haegeman2012} to compute dispersion relations. We consider looking for a quadratic approximation to the Hamiltonian of the form
\begin{equation}
\label{eq:bosonicHamilt}
\begin{split}
    H_B = &E_0+\sum_{ix\mu\nu}\epsilon_{\mu\nu}^{(x)}a_{\mu}^{\dagger(i)}a_\nu^{(i+x)}\\ &+ \sum_{\substack{i\mu\nu \\x>0}}\Delta_{\mu\nu}^{(x)}a_{\mu}^{\dagger(i)}a_{\nu}^{\dagger(i+x)}+\overline{\Delta}_{\mu\nu}^{(x)}a_{\mu}^{(i)}a_{\nu}^{(i+x)}.
\end{split}
\end{equation}
with additional terms at third order or higher. Normally we would also obtain first order terms but we assume the expansion is performed around a local minimum of the Hamiltonian on the MPS manifold, so the gradient vanishes. The idea is to relate the coefficients in the expansion to the matrix elements of the physical Hamiltonian in the basis of tangent space vectors

\begin{align}
    E_0 &= \bra{\Psi} H\ket{\Psi}, \\
    \epsilon_{\mu\nu}^{(ij)} &= \bra{\partial_\mu^{(i)}\Psi} H \ket{\partial_\nu^{(j)} \Psi} - E_0\delta_{\mu\nu}\delta_{ij}, \\
    \Delta_{\mu\nu}^{'(ij)} &= \bra{\partial_\mu^{(i)}\partial_\nu^{(j)}\Psi} H\ket{\Psi}.
\end{align}
Due to the Gram matrix not being identity in the second tangent space, it can be shown that these coefficients correspond to the quadratic terms in the operator $H_B' = \mathcal{G}H_B$ rather than the one we want $H_B$. The annihilation part remains unaltered, so $\overline{\Delta} = \Delta^{'*}$, but using $\Delta'$ for the creation part of the bosonic Hamiltonian leads to too many excitations being produced, since the vectors $\ket{\partial_\mu^{(i)}\partial_\nu^{(j)}\Psi}$ are not linearly independent. To obtain the correct $\Delta$ one needs to apply the inverse of the Gram matrix, which is explained further in the Supplementary Material.

The second method builds the Hamiltonian $H_B$ directly by looking at how the terms in the Hamiltonian interact with the structure of the local tensors of the MPS. To see how this works, consider some matrix $M$ that acts on both the physical and the bond indices of the output of tensor $A$. The pull-through equation consists of writing this tensor as a linear combination of a local derivative and a term with a matrix on the opposite side of $A$. This can be expressed diagrammatically as
\begin{equation}
\begin{split} &
\begin{tikzpicture}[baseline={([yshift=-0.6ex]current bounding box.center)}]
\Vertex[x = 0, y = 0.25, size = 0.7]{M1}
\Vertex[x = 0, y = -0.25, size = 0]{M2}
\Vertex[x = -0.1, y = -0.1, size = 0]{M3}
\Vertex[label=$M$, shape=rectangle, size=0.7, x = 0, y = 0,style={minimum height=1.2cm},RGB,color={253,192,134}]{M}
\Vertex[label=$A$,size=0.7, x = 1, y = 0.25]{A}
\Vertex[x = -1,y = 0.25,style = {color=white}]{B}
\Vertex[x = -1,y = -0.25,style = {color=white}]{C}
\Vertex[x = 2,y = 0.25,style = {color=white}]{D}
\Edge(A)(M1)
\Edge[bend = 40](A)(M3)
\Edge(M1)(B)
\Edge(M2)(C)
\Edge(A)(D)
\end{tikzpicture} = \begin{tikzpicture}[baseline={([yshift=-0.6ex]current bounding box.center)}]
\Vertex[x = 0, y = 0.25, size = 0.7]{M1}
\Vertex[x = 0, y = -0.25, size = 0]{M2}
\Vertex[x = -0.1, y = -0.1, size = 0]{M3}
\Vertex[label=$U^\dagger M$, shape=rectangle, size=0.9, x = 0, y = 0,style={minimum height=1.2cm},RGB,color={253,192,134}]{M}
\Vertex[label=$A$,size=0.7, x = 1, y = 0.25]{A}
\Vertex[x = 2,y = 0.25,style = {color=white}]{D}
\Vertex[x = -2.5,y = 0.25,style = {color=white}]{E}
\Vertex[x = -2.5,y = -0.25,style = {color=white}]{F}
\Vertex[label=$U$, shape=rectangle, size=0.7, x = -1.5, y = 0, style={minimum height=1.2cm}]{U}
\Edge(A)(M1)
\Edge[bend = 40](A)(M3)
\Edge(M1)(E)
\Edge(M2)(F)
\Edge(A)(D)
\node at (-0.83, -0.45) {\tiny $i \neq 0 $};
\end{tikzpicture}  \\
&\hspace{1.5cm}+ \begin{tikzpicture}[baseline={([yshift=-0.6ex]current bounding box.center)}]
\Vertex[x = 0, y = 0.25, size = 0.7]{M1}
\Vertex[x = 0, y = -0.25, size = 0]{M2}
\Vertex[x = -0.1, y = -0.1, size = 0]{M3}
\Vertex[x = 0.1, y = -0.1, size = 0]{M4}
\Vertex[label=$M$, shape=rectangle, size=0.7, x = 0, y = 0,style={minimum height=1.2cm},RGB,color={253,192,134}]{M}
\Vertex[label=$A$,size=0.7, x = 1, y = 0.25]{A}
\Vertex[label=$A$,size=0.7, x = -2, y = 0.25]{A3}
\Vertex[label=$A^\dagger$,size=0.7, x = -1, y = 0.25]{A2}
\Vertex[x = 2,y = 0.25,style = {color=white}]{D}
\Vertex[x = -3,y = 0.25,style = {color=white}]{B}
\Vertex[x = -3, y = -0.25 ,style = {color=white}]{C}
\Edge(A)(M1)
\Edge[bend = 40](A)(M3)
\Edge[bend = -40](A2)(M4)
\Edge(A)(D)
\Edge(A2)(M1)
\Edge(A3)(A2)
\Edge(A3)(B)
\Edge[bend = 40](A3)(C)
\end{tikzpicture}.
\end{split}
\end{equation}
The first term on the RHS is a local derivative term parameterized by the 3-tensor $x = (1-\ket{0}\bra{0}_p)U^\dagger M A \sqrt{\Gamma}$ (compare with Eq.~\eqref{eq:derivative}). In the second term $M$ has been 'pulled through' to the other side of the local tensor $A$. Note that in a MPS, this equation can be applied recursively to move $M$ to the right end of the chain. The properties of the transfer matrix governing the pull-through map $M \to A^\dagger M A$ guarantee that all terms decay exponentially quickly, except for the identity. The same principle holds for pulling matrices through derivative tensors, with the details found in the Supplementary Material. If we apply this procedure to initial operators $h$ acting in the physical space, we end up with a mapping similar to the Holstein-Primakoff transformation from the set of single-site physical operators $h$ to the bosonic space. The only problem with this approach is that it does not automatically remove the zero-norm two-particle states, so these must still be projected out after constructing the quasi-particle Hamiltonian.

\begin{figure}[t!]
    \centering
    \includegraphics[width=0.47\textwidth]{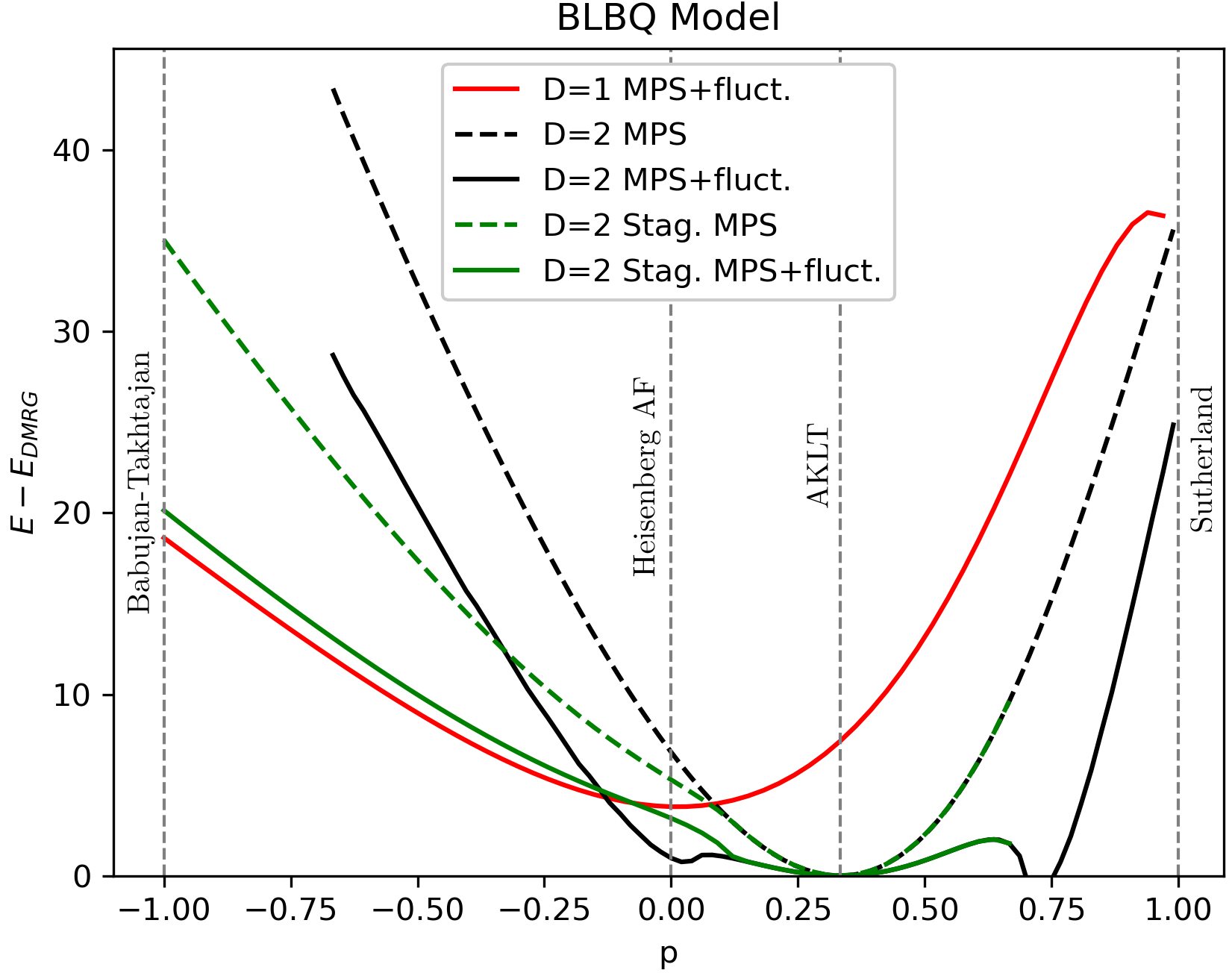}
    \caption{Ground state energy approximations for the BLBQ model of Eq.~\eqref{eq:BLBQ} in the Haldane phase corresponding to the parameter $p$ in the range -1 to 1. The energy surplus is compared to a $\chi = 50$ DMRG. The $D=1$ and the $D=2$ Stag. results are obtained by applying a flip on every other site. Dashed lines mark single MPS results and bold lines include fluctuation corrections.}
    \label{fig:label}
\end{figure}

\textit{Fluctuation correction}---to illustrate the power of our approach we consider an example task of calculating the fluctuation corrections to the ground state energy. This is equivalent to solving the following minimisation equation
\begin{equation}
    E = \min_{\phi} \frac{[\phi| \mathcal{G}H_B|\phi]}{[\phi| \mathcal{G}|\phi]},
\end{equation}
where $|\phi]$ is a bosonic state. This leads to the generalized eigenvalue equation
\begin{equation}
    \mathcal{G}H_B |\phi] = E \mathcal{G} |\phi], \hspace{1cm} \mathcal{G}|\phi] \neq 0
\end{equation}
which is complicated by the fact that $\mathcal{G}$ is not invertible due to the zero-norm states discussed previously. The way to proceed is to construct the pseudo-inverse $\mathcal{G}^+$ and apply it to both sides. Then we see that it is sufficient to solve the regular eigenvalue equation for the projected Hamiltonian $\mathcal{G}^+\mathcal{G} \mathcal{H}$ which does not produce any zero-norm pairs of excitations. We truncate this operator at quadratic order and use the known techniques to calculate the fluctuation correction to the ground state of a bosonic system with anomalous terms \cite{Derezinski2017}.

\textit{Results}---We test the algorithm described above for computing fluctuation corrections of the following set of parameterized spin-$1$ Hamiltonians
\begin{equation}
\label{eq:BLBQ}
    \mathcal{H} = \sum_i  \Vec{S}_i\cdot \Vec{S}_{i+1} + p \left( \Vec{S}_i\cdot \Vec{S}_{i+1}\right)^2,
\end{equation}
often called the isotropic bilinear-biquadratic (BLBQ) Heisenberg model \cite{Fath1995, Rakov2022, Lauchli2006}. This is known to be in the Haldane phase for $p$ between $-1$ and $1$, which includes the Heisenberg antiferromagnet at $p =0$ and the AKLT model at $p = 1/3$. The reason for this choice is the ground state of the AKLT model is exactly an MPS with $D=2$. Despite no longer being the exact ground state, it remains a saddle point of the TDVP equations for a wider range of $\theta$ values in the same phase, so it forms a good reference state. Its transfer matrix is also very simple, allowing us to gain analytical insight into the structure of Gram operators. This is detailed in the Supplementary Material. The fluctuation corrections to the ground state energy around the AKLT state (as well as a D=1 Néel state for comparison) at various values of the parameter $p$ are shown in Fig.~\ref{fig:label}. The fluctuation corrections are seen to give a significant improvement in the region around the AKLT $p=1/3$. However, the $D=2$ MPS manifold predicts dimerization at $p\sim 0.1$, far from the theoretical value $p=-1$. This is a consequence of the mean-field nature of the approach, and we should only expect the true phase transition to appear once we have included interaction terms in the effective Hamiltonian Eq.~\eqref{eq:bosonicHamilt}.

\begin{figure}[t!]
    \centering
    \includegraphics[width=0.5\textwidth]{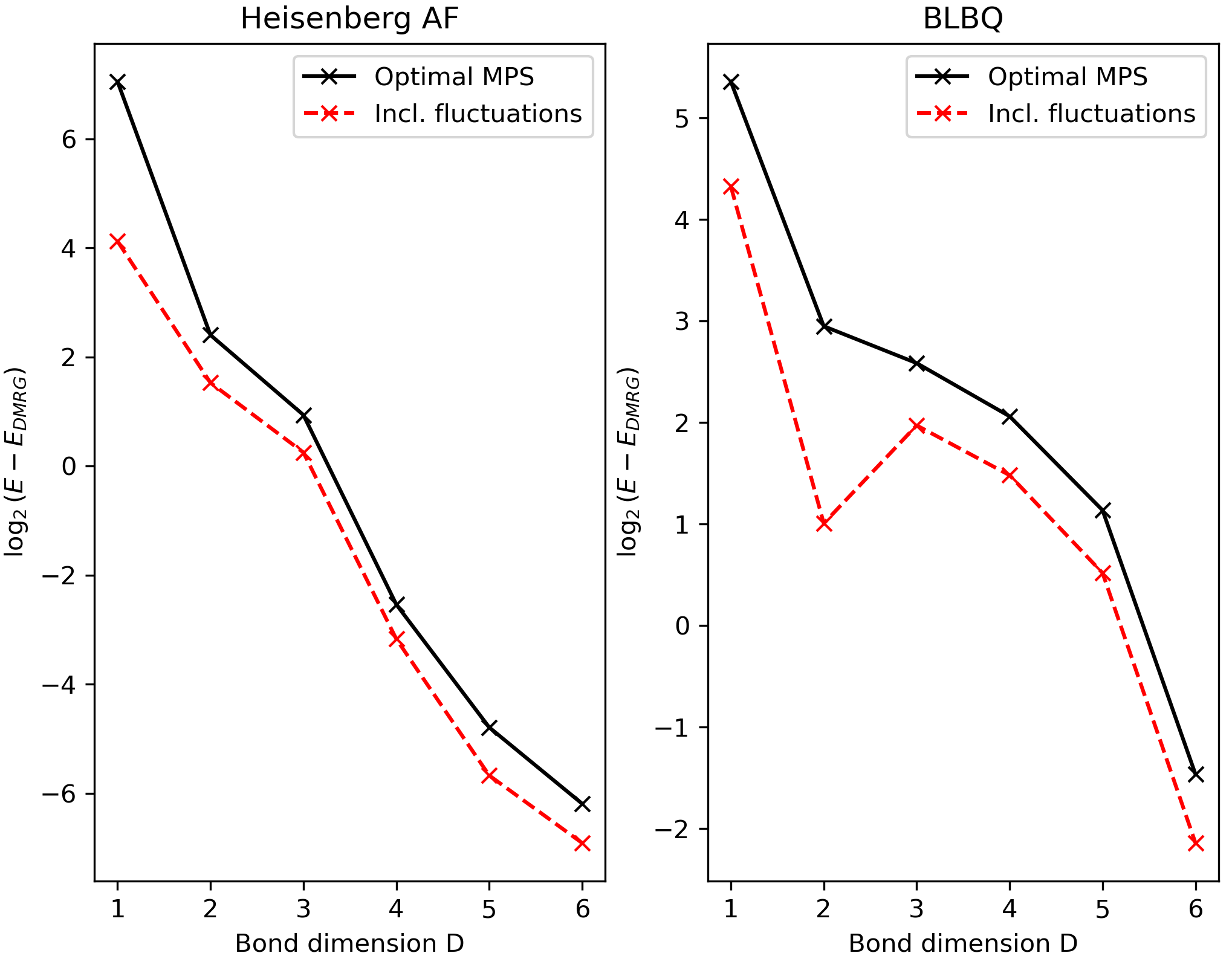}
    \caption{Energy corrections obtained for various bond dimensions with and without zero-point fluctuation energy, compared to a $\chi = 50$ DMRG calculation. Left panel is for the $S=1/2$ Heisenberg antiferromagnet with an applied staggered magnetic field. Right panel is for the $S=1$ BLBQ model with a particular choice of $p = p_{AKLT}+0.3$.}
    \label{fig:enter-label}
\end{figure}

We also check how the performance of the algorithm is affected by the choice of bond dimension. For this purpose, we consider both the spin-1 BLBQ model described above with $\Delta p = 0.3 = p-p_{AKLT}$, as well as the spin-1/2 Heisenberg model with an applied staggering field $h = 0.2$ to remove ground state degeneracy. The first thing we note is the free particle fluctuation correction constitutes the largest fraction of the remaining energy at the lowest bond dimension that allows for a qualitatively correct reference state. For the Heisenberg antiferromagnet this is the $D=1$ Néel state, while the BLBQ model is in the Haldane phase, which is qualitatively similar to the $D=2$ AKLT state. As the bond dimension is increased beyond this point, we find that the fluctuation correction approaches a finite fraction of the gap, which is represented by a nearly constant distance between the graphs in Fig.~\ref{fig:enter-label}. This suggests a perturbative treatment of the interactions may be easiest by working on the lowest bond dimension manifold that is able to qualitatively capture the entanglement structure of the true ground state.

\textit{Discussion}---We presented a method for constructing a bosonic representation of tensor fluctuations around an MPS reference point. This can be understood as a generalization of the Holstein-Primakoff transformation for treating complex systems where primary excitations driving the low-energy physics are delocalised over multiples sites. We illustrate the method by constructing the effective quadratic bosonic Hamiltonian of the bilinear-biquadratic Heisenberg model inside the Haldane phase, where the bond dimension 2 AKLT ground state is a good reference. Despite neglecting interactions and truncating the Gram operator at first non-trivial order, we showed that the fluctuation corrections due to the pairing of AKLT spinons accounts for a large portion of the distance to the energy of the true ground state. Due to the simplicity associated with free boson theories and our method for constructing bosonic equivalents for arbitrary spin operators, a large number of quantities of interest could be derived, such as order parameters, diffusion coefficients and critical exponents. We do not discuss topological excitations or non-translationally invariant reference states here, but we expect they can be included in a similar fashion, at the expense of dealing with more complicated Gram operators. The method could also be further developed to account for interactions at mean-field level self-consistently, which can lead to large corrections if the local occupations of the bosonic modes are large.

The effect of both the bosonic Hamiltonian and the Gram operator could be captured by a path integral over the MPS manifold, which has been previously proposed \cite{Green2016}. In this formulation, the Gram operator appears as a local 4'th order modification to the bosonic Berry phase, which has been shown here and in previous work \cite{Leontica2024} to contribute meaningfully at the free boson level of approximation. A more thorough investigation of this connection and its implications for the possibility of constructing path integrals over entangled phase spaces are left for future work. Combined with renormalization group methods, this approach could lead to simple effective field theories of phase transitions beyond those accessible by the Landau-Ginzburg mean-field approach.

Finally, we remark that our work could also prove a very effective theoretical tool for studying exotic materials, by systematically constructing tensor approximations to local excitations along with their interaction Hamiltonian. The fluctuation corrections obtained from the Gaussian approximation are only meant as an illustration of how the techniques described could be used in practice. Constructing the effective Hamiltonian to higher orders is straight-forward in principle by following the pull-through equations detailed in the Supplemental Material, but dealing with the removal of zero-norm states to higher order will require a more thorough investigation into the Gram operator. It is nevertheless exciting to speculate that the principles outlined here could eventually be used to automate the identification of key low-energy physics (in 1D and possibly beyond) and to propose effective field theories for their excitations.

\vspace{0.5cm}
\textit{Acknowledgments}---S.L. acknowledges support from UCL’s Graduate Research Scholarship and Overseas Research Scholarship. The authors acknowledge support from the EPSRC under grants EP/W026872/1, EP/S005021/1.

\bibliography{bibliography}

\onecolumngrid

\section{Free boson Hamiltonian and the Gram operator}
\label{sec:bosonicmapping}

In this section, we provide an overview of how to obtain the free-boson Hamiltonian for the tensor fluctuations and how to iteratively construct the Gram operator. The construction relies on interpreting MPS states with derivatives $\partial_\mu^{(i)}$ as bosonic states at site $i$ of flavour $\mu$ on top of an MPS reference vacuum. Then we can proceed by looking for the best quadratic bosonic Hamiltonian that approximates the matrix elements of the true spin Hamiltonian in the basis of MPS with local derivatives. We focus here only on the machinery useful for constructing the quadratic approximation used in the main text. A full exposition of how to obtain a more general mapping between a physical Hamiltonian and a bosonic theory of MPS perturbations is deferred to Sec.~\ref{sec:holsteinPrimakoff} below.

If $H$ denotes the physical spin Hamiltonian and $H_B$ is it's bosonic equivalent then we want to find an expansion

\begin{equation}
    H_B' = E_0 + \sum_{i,\mu}\left( h_\mu^{(i)}a_\mu^{(i)\dagger} + h^{(i)*}_\mu a_\mu^{(i)}\right)+\sum_{ix\mu\nu}\epsilon_{\mu\nu}^{(x)}a_{\mu}^{\dagger(i)}a_\nu^{(i+x)} + \sum_{\substack{i\mu\nu \\x>0}}\Delta_{\mu\nu}^{\prime (x)}a_{\mu}^{\dagger(i)}a_{\nu}^{\dagger(i+x)}+\overline{\Delta}_{\mu\nu}^{\prime (x)}a_{\mu}^{(i)}a_{\nu}^{(i+x)} + \mathcal{O}(a^3),
\end{equation}
which reproduces the matrix elements of the true Hamiltonian under the mapping
\begin{equation}
\bra{\prod_{\mu,i}\partial_\mu^{(i)}\Psi} H\ket{\prod_{\nu,j}\partial_\nu^{(j)}\Psi} = [GS|\prod_{\mu,i}a_\mu^{(i)} H_B'\prod_{\nu,j}a_\nu^{\dagger(j)}|GS],
\end{equation}
where the $|\cdot ]$ notation is used to distinguish states in the bosonic virtual space from those in the physical Hilbert space, with a more concrete definition given in Sec.\ref{sec:holsteinPrimakoff}. In principle this equivalence can be carried out iteratively to all orders, but to find the best quadratic approximation we only want to match matrix elements with up to 2 derivatives in total. A simple inspection reveals that
\begin{align}
    E_0 &= \bra{\Psi} H\ket{\Psi}, \\
    h_\mu^{(i)} &= \bra{\partial_\mu^{(i)}\Psi} H\ket{\Psi}\\
    \epsilon_{\mu\nu}^{(ij)} &= \bra{\partial_\mu^{(i)}\Psi} H\ket{\partial_\nu^{(j)} \Psi} - E_0\delta_{\mu\nu}\delta_{ij}, \\
    \Delta_{\mu\nu}^{'(ij)} &= \bra{\partial_\mu^{(i)}\partial_\nu^{(j)}\Psi} H\ket{\Psi},
\end{align}
is the correct choice. The reason this is expected to be a good approximation is that the influence of placing a derivative in the MPS only affects the many-body state locally, as shown by the transfer matrix argument in the main text. This means the bosonic Hamiltonian is also expected to be local, with higher power contributions being exponentially suppressed. This is also corroborated by the 'Holstein-Primakoff' representation described in Sec.~\ref{sec:holsteinPrimakoff}. Let us now attempt to formulate the ground state eigenvalue problem of the spin Hamiltonian $H$ in the bosonic picture. We parameterize the spin Hilbert space using the overcomplete basis of derivatives to the MPS reference state as
\begin{equation}
    \ket{\Psi(x)} = \ket{\Psi} + \sum_{i\mu} x_\mu^{(i)} \ket{\partial_\mu^{(i)}\Psi} + \sum_{ij\mu\nu}x_{\mu\nu}^{(ij)}\ket{\partial_\mu^{(i)}\partial_\nu^{(j)}\Psi} + \ldots,
\end{equation}
and formulate the variational principle for the ground state energy
\begin{equation}
\label{eq:varprinc}
    E_{\operatorname{GS}} = \min_x \frac{\bra{\Psi(x)} H \ket{\Psi(x)}}{\bra{\Psi(x)} \ket{\Psi(x)}}.
\end{equation}
On the basis of the mapping we found above we see that we can replace $\bra{\Psi(x)} H \ket{\Psi(x)} = [\phi(x)|H_B'|\phi(x)]$ where $|\phi(x)]$ is the bosonic state obtained by replacing derivatives by bosonic excitations. The only issue we face now is the state normalization, i.e. the denominator in Eq.~\eqref{eq:varprinc} above. It is clear that $\bra{\Psi(x)} \ket{\Psi(x)} \neq [\phi(x)| \phi(x)]$ in general, but we can play the same trick as before and look for a bosonic operator $\mathcal{G}$ such that
\begin{equation}
\bra{\prod_{\mu,i}\partial_\mu^{(i)}\Psi} \ket{\prod_{\nu,j}\partial_\nu^{(j)}\Psi} = [GS|\prod_{\mu,i}a_\mu^{(i)} \mathcal{G} \prod_{\nu,j}a_\nu^{\dagger(j)}|GS],
\end{equation}
for all elements in the derivative basis. Since this is equivalent to finding the bosonic representative of the simple Hamiltonian $H=1$ we also expect $\mathcal{G}$ to be well approximated by local few-body operators. However, for reasons that will become clear later, a consistent calculation at second order in the Hamiltonian requires keeping terms in the Gram operator up to fourth order with 2 creation and 2 annihilation operators. Due to our choice of basis in the first tangent space, overlaps between states with a single excitation are exactly equal to their bosonic counterpart, such that $\mathcal{G} = 1 + \mathcal{O}(a^3)$. To find its expansion further let us take a closer look at states in the second tangent space $\ket{\partial_\mu^{(i)}\partial_\nu^{(j)}\Psi}$. Matters are complicated by the fact that there is non-zero overlap between the second and first tangent space when $i$ and $j$ are close to each other. The contributions at third order in the Gram operator should then account for the non-zero overlaps

\begin{equation}
    M_{\kappa\mu\nu}^{(x)} = \bra{\partial_\kappa^{(0)}\Psi}\ket{\partial_\mu^{(0)}\partial_\nu^{(x)}\Psi},
\end{equation}
where $i<j$. We can deal with this issue by interpreting the bosonic excitations as covariant derivatives rather than simple partial derivatives. The connection between these two states in the second tangent space is then
\begin{equation}
    \ket{\nabla_\mu^{(i)} \nabla_\nu^{(j)} \Psi} = \ket{\partial_\mu^{(i)} \partial_\nu^{(j)} \Psi} - \sum_\kappa M_{\kappa\mu\nu}^{(j-i)} \ket{\partial_\kappa^{(i)}\Psi},
\end{equation}
where these are now orthogonal to the entire first tangent space. With this reinterpretation of the bosonic modes we have to correct the expression for $\Delta'$ as
\begin{equation}
    \Delta^{\prime (ij)}_{\mu\nu} = \bra{\nabla_\mu^{(i)} \nabla_\nu^{(j)} \Psi} H \ket{\Psi} = \bra{\partial_\mu^{(i)} \partial_\nu^{(j)} \Psi} H \ket{\Psi} - \sum_\kappa M_{\kappa\mu\nu}^{(j-i)} h_\kappa^{(i)},
\end{equation}
but in practice this does not affect fluctuation calculations as we always expand around saddle points $h_\kappa^{(i)} = 0$. Similar redefinitions should be possible in higher tangent spaces in order to make them independent and restrict our attention only to terms in the Gram operator that have an equal number of creation and annihilation operators. After this redefinition we have the generic normal ordered form
\begin{equation}
\mathcal{G} = 1 +  \sum_{\substack{\mu\nu\alpha\beta,i\\x,y>0}}g_{\mu\nu\alpha\beta}^{(xy)}a_\mu^{(i)\dagger} a_\nu^{(i+x)\dagger} a_\alpha^{(i)} a_\beta^{(i+y)} + \ldots.
\end{equation}
The coefficients $g$ are related to the the Gram matrix $\Tilde{G}$ in the second tangent space via
\begin{equation}
    g^{(xy)}_{\mu\nu,\alpha\beta} = \tilde{G}^{(xy)}_{\mu\nu,\alpha\beta} - \delta_{xy}\delta_{\mu\alpha}\delta_{\nu\beta},
\end{equation}
where
\begin{align}
    &\tilde{G}^{(xy)}_{\mu\nu,\alpha\beta} = \bra{\nabla_\mu^{(0)}\nabla_\nu^{(x)}\Psi} \ket{\nabla_\alpha^{(0)}\nabla_\beta^{(y)}\Psi} = G^{(xy)}_{\mu\nu,\alpha\beta} - \sum_\kappa M^{(j)*}_{\kappa\mu\nu}M^{(l)}_{\kappa\alpha\beta}, \\
    &G^{(xy)}_{\mu\nu,\alpha\beta} = \bra{\partial_\mu^{(0)}\partial_\nu^{(x)}\Psi} \ket{\partial_\alpha^{(0)}\partial_\beta^{(y)}\Psi}.
\end{align}

The derivative basis is an overcomplete basis for the Hilbert space, as could be easily seen by noting that that the number of possible tensors per site is $1+D^2(d-1)$, which is larger than the size of the physical Hilbert space of dimension $d$ when $D>1$. This is compensated by a reduction in the number of physical states in higher tangent spaces, or in our case in the non-invertibility of the Gram operator. In Sec.~\ref{sec:AKLTgram} we show this explicitly by computing the eigenvalues of the Gram matrix $\tilde{G}$ exactly for the AKLT reference state.

If we use this to express the state norm in the bosonic basis and formulate the variational principle over bosonic states we see that the ground state is the solution to the generalized eigenvalue problem
\begin{equation}
    H_B'|\phi] = E_{\operatorname{GS}}\mathcal{G} |\phi].
\end{equation}

Applying the Moore-Penrose inverse $\mathcal{G}^+$ to both sides this can be transformed to the eigenvalue problem of $H_B = \mathcal{G}^+ H_B'$. If we can find the quadratic approximation of $H_B$, then we can use the known techniques of working with Gaussian bosonic states to compute its ground state energy. The ansatz for this Hamiltonian is similar to the previous

\begin{equation}
    H_B = E_0 +\sum_{ix\mu\nu}\epsilon_{\mu\nu}^{(x)}a_{\mu}^{\dagger(i)}a_\nu^{(i+x)} + \sum_{\substack{i\mu\nu \\x>0}}\Delta_{\mu\nu}^{ (x)}a_{\mu}^{\dagger(i)}a_{\nu}^{\dagger(i+x)}+\overline{\Delta}_{\mu\nu}^{(x)}a_{\mu}^{(i)}a_{\nu}^{(i+x)} + \mathcal{O}(a^3),
\end{equation}
where $E_0$, $\epsilon$ and $\overline{\Delta} = \overline{\Delta}'$, but $\Delta$ is related to $\Delta'$ according to
\begin{equation}
    \Delta_{\mu\nu}^{(j)} = \sum_{\alpha\beta, l>0} \tilde{G}^{(jl)}_{\mu\nu,\alpha\beta} \Delta^{\prime (l)}_{\alpha\beta}.
\end{equation}

Since the Hamiltonian $H_B'$ is hermitian by construction, such that $\overline{\Delta}'$ is the complex conjugate of $\Delta'$, the Hamiltonian $H_B$ whose lowest eigenvalue gives the ground state energy will not be Hermitian in general. The coefficients in the expansion of $H_B'$ can be easily calculated as tensor network contractions which are efficient in 1D. The most computationally expensive step is calculating the inverse of $\tilde{G}$, which is an $LD^2(d-1) \times LD^2(d-1)$ matrix, where the length $L$ can be taken on the order of a few correlation lengths.

\section{The AKLT Gram operator}
\label{sec:AKLTgram}

In this chapter we will give an overview of the qualitative features of the second tangent space Gram matrix by looking at the simplest non-trivial example, the AKLT ground state. This is a bond dimension $D=2$ MPS and it is particularly simple because its transfer matrix has a single three-fold degenerate sub-leading eigenvalue $\lambda = -1/3$. This means the matrices introduced in the previous section have a very simple space dependence given by the formulae
\begin{align}
G_{\mu\nu,\alpha\beta}^{(ii)} &= \delta_{\mu\alpha}\delta_{\nu\beta}\left(1-\lambda^{i-1}\right)+\lambda^{i-1} G_{\mu\nu,\alpha\beta}^{(11)},\\
G_{\mu\nu,\alpha\beta}^{(ij)} &= \lambda^{j-2} G_{\mu\nu,\alpha\beta}^{(12)}, \hspace{3pt} j>i,\\
M_{\kappa\mu\nu}^{(i)} &= \lambda^{i-1}M_{\kappa\mu\nu}^{(1)}.
\end{align}

We can then use these to construct $\tilde{G}$ in terms of the 3 basic matrices $L = G^{(11)}$, $J = G^{(12)}$ and $M = M^{(1)}$. If $\Vec{v}$ is an eigenvector with eigenvalue $E$ and $\vec{v}_l$ are the sub-vectors corresponding to sites $l$ from $1$ to $N$ then we can show the following second order recurrence relation must hold

\begin{equation}
    A_j \vec{v}_{j+2} - B_j \vec{v}_{j+1} + C_j \Vec{v}_j = 0,
\end{equation}
where the coefficient matrices are given by
\begin{equation}
    \begin{split}
        A_j &= \lambda^j\left(J-\lambda L+\lambda\right) + (E-1), \\
        B_j &=  \lambda^j (J+J^\dagger) + (1+\lambda)\left(\lambda^j-\lambda^j L + (E-1)\right), \\
        C_j &= \lambda^{j-1}(J^\dagger - \lambda L + \lambda I) + \lambda (E-1),
    \end{split}
\end{equation}
and the $j$ can be any integer, with the understanding that $\vec{v}_j = 0$ for all $j\leq 0$. Note that the matrix $M$ does not enter the expressions. At the present time we are not aware if such recurrence relations can be found for more general transfer matrices, which is an important question as it would vastly simplify the task of inverting the Gram matrix.

Consider the following ansatz as a possible solution to the recurrence
\begin{equation}
\begin{split}
    \vec{v}_j &= 0, \text{when } j<y, \\
    \vec{v}_y &= \vec{v}, \\
    \vec{v}_j &= \sum_{s=0}^\infty \lambda^{s(j-y)} \vec{\alpha}_s, \text{when } j > y.
\end{split}
\end{equation}

For the starting term of any sequence we have to take
\begin{equation}
    A_{y-2} \vec v = 0,
\end{equation}
which means the starting vector must be a solution to
\begin{equation}
    \left[\lambda^{y-2}(J-\lambda L + \lambda) + (E-1)\right] \vec{v} = 0,
\end{equation}
and therefore an eigenvector of $J-\lambda L$. The remaining equations tell us about the vector tail. The eigenvalues of $J-\lambda L$ turn out to be $0$ or $1-\lambda$. Then the two possible eigenvalues corresponding to a vector with starting position $y$ are
\begin{equation}
\begin{split}
    E^{(y)}_{+} = 1-\lambda^{y-2}, \\
    E^{(y)}_{-} = 1-\lambda^{y-1}.
\end{split}
\end{equation}

Then we see that the Gram matrix indeed has a number of null eigenvectors, corresponding to starting positions $y=1,2$. As expected from the MPS locality argument we see that in the limit $y \gg -\log \abs{\lambda}$ the derivatives become independent.

\section{The 'Holstein-Primakoff' mapping for MPS}
\label{sec:holsteinPrimakoff}

In this section, we will present a method for obtaining a Holstein-Primakoff type expression for the action of local spin operators as an expansion in MPS derivatives. The standard Holstein-Primakoff is an exact mapping between operators acting on spin $S$ particles and the creation and annihilation operators $(a,a^\dagger)$ of a bosonic mode. The relation is usually stated as
\begin{equation}
\label{eq:HP}
S_{+}=\hbar \sqrt{2 s} \sqrt{1-\frac{a^{\dagger} a}{2 s}} a, \quad S_{-}=\hbar \sqrt{2 s} a^{\dagger} \sqrt{1-\frac{a^{\dagger} a}{2 s}}, \quad S_z=\hbar\left(s-a^{\dagger} a\right),
\end{equation}
and it forms a good starting point for a semi-classical treatment of spin-waves at large $S$ by Taylor expanding the square root. The mapping is performed locally and it associates one bosonic mode per site. In our case, the bosonic excitations created by the action of this operator will bear the physical interpretation of perturbations to the tensor structure, with one mode per site corresponding to each orthogonal direction in which the tensor can be modified, or equivalently to each derivative tensor $B_\mu$ defined in the main text as
\begin{equation}
    \begin{tikzpicture}[baseline={([yshift=2ex]current bounding box.center)}]
        \Vertex[label = $B_\mu$,size = 0.7]{A} \Vertex[x = 1,style = {color=white}]{B} \Vertex[x=-1,style = {color=white}]{C} \Vertex[y=-1,style = {color=white}]{D}
        \Edge(A)(B)
        \Edge(A)(C)
        \Edge(A)(D)
    \end{tikzpicture} = 
    \begin{tikzpicture}[baseline={([yshift=1ex]current bounding box.center)}]
        \Vertex[label = $U$,shape = rectangle,size = 0.7]{A} \Vertex[label = $e_\mu$,x = 1,RGB,color = {190, 221, 186}]{B} \Vertex[x=-1,style = {color=white}]{C} \Vertex[y=-1,style = {color=white}]{D} 
        \Vertex[label = $\Gamma^{-\frac{1}{2}}$,size = 0.7,x = 2,RGB,color={253,192,134}]{G} \Vertex[x=3, style = {color=white}]{H}
        \Edge[Direct](B)(A)
        \Edge[Direct](A)(C)
        \Edge[Direct](A)(D)
        \Edge(G)(B)
        \Edge[bend = -90,Direct](B)(A)
        \Edge(H)(G)
    \end{tikzpicture},
\end{equation}
where $U$ is the reference unitary, $\Gamma$ is the right environment of the reference state and $e_\mu$ is a unit tensor with $1$ in some position and $0$ throughout the rest. The additional restriction that $e_\mu$ must not be $0$ on the top index means we have a total of $D^2(d-1)$ bosonic modes per site.

The interpretation of such fluctuations as bosonic excitations that we construct bears some resemblance to the Holstein-Primakoff mapping, but there are also important differences. Our method allows us to express expectation values of observables in states described via tensor networks as expectation values of bosonic operators in associated bosonic states. It does not provide an exact mapping between states and operators in the same sense as Eq.~\eqref{eq:HP}. Instead, we will find that, following the procedure outlined below, one can map expectation values such as
\begin{equation}
\bra{\partial^{(i_1)}_{\mu_1}\partial^{(i_2)}_{\mu_2}\cdots \Psi} \cdots \ket{\partial^{(j_1)}_{\nu_1}\partial^{(j_2)}_{\nu_2}\cdots \Psi} = [GS|\left(a^{(i_1)}_{\mu_1} a^{(i_2)}_{\mu_2} \cdots\right) \mathcal{G}\cdots \left(a^{(j_1)\dagger}_{\nu_1}a^{(j_2)\dagger}_{\nu_2}\cdots\right)|GS],
\end{equation}
where the derivatives on the LHS have a one to one correspondence to the excitations created on the RHS and the Gram operator is the one defined in Sec.~\ref{sec:bosonicmapping}. The $|\cdot]$ notation is used to distinguish the bosonic space from the physical Hilbert space, and will acquire a more formal definition below. We find that all local operators acting on the physical Hilbert space have a quasi-local bosonic representation that satisfies the expression above for all possible derivative configurations, and we give a recipe for how to construct it efficiently.

The primary tool we use in our construction is the pull-through equation. If we consider an operator $M$ acting on both the left bond index and the physical index of the reference tensor $A$ at some site, we see that we can split this as
\begin{equation}
\begin{split}
        \begin{tikzpicture}[baseline={([yshift=-0.6ex]current bounding box.center)}]
\Vertex[x = 0, y = 0.25, size = 0.7]{M1}
\Vertex[x = 0, y = -0.25, size = 0]{M2}
\Vertex[x = -0.1, y = -0.1, size = 0]{M3}
\Vertex[label=$M$, shape=rectangle, size=0.7, x = 0, y = 0,style={minimum height=1.2cm},RGB,color={253,192,134}]{M}
\Vertex[label=$A$,size=0.7, x = 1, y = 0.25]{A}
\Vertex[x = -1,y = 0.25,style = {color=white}]{B}
\Vertex[x = -1,y = -0.25,style = {color=white}]{C}
\Vertex[x = 2,y = 0.25,style = {color=white}]{D}
\Edge(A)(M1)
\Edge[bend = 40](A)(M3)
\Edge(M1)(B)
\Edge(M2)(C)
\Edge(A)(D)
\end{tikzpicture} = 
\begin{tikzpicture}[baseline={([yshift=-0.6ex]current bounding box.center)}]
\Vertex[x = 0, y = 0.25, size = 0.7]{M1}
\Vertex[x = 0, y = -0.25, size = 0]{M2}
\Vertex[x = -0.1, y = -0.1, size = 0]{M3}
\Vertex[label=$M$, shape=rectangle, size=0.7, x = 0, y = 0,style={minimum height=1.2cm},RGB,color={253,192,134}]{M}
\Vertex[label=$A$,size=0.7, x = 1, y = 0.25]{A}
\Vertex[x = 2,y = 0.25,style = {color=white}]{D}
\Vertex[x = -3.5,y = 0.25,style = {color=white}]{E}
\Vertex[x = -3.5,y = -0.25,style = {color=white}]{F}
\Vertex[label=$U$, shape=rectangle, size=0.7, x = -2.5, y = 0, style={minimum height=1.2cm}]{U}
\Vertex[label=$U^\dagger$, shape=rectangle, size=0.7, x = -1, y = 0, style={minimum height=1.2cm}]{U}
\Edge(A)(M1)
\Edge[bend = 40](A)(M3)
\Edge(M1)(E)
\Edge(M2)(F)
\Edge(A)(D)
\node at (-1, -1.25) {Split to $\ket{0}\bra{0}$ and $1-\ket{0}\bra{0}$};
\draw[->] (-1.75, -1) -- (-1.75, -0.5);
\end{tikzpicture}\\
=\begin{tikzpicture}[baseline={([yshift=-0.6ex]current bounding box.center)}]
\Vertex[x = 0, y = 0.25, size = 0.7]{M1}
\Vertex[x = 0, y = -0.25, size = 0]{M2}
\Vertex[x = -0.1, y = -0.1, size = 0]{M3}
\Vertex[label=$U^\dagger M$, shape=rectangle, size=0.9, x = 0, y = 0,style={minimum height=1.2cm},RGB,color={253,192,134}]{M}
\Vertex[label=$A$,size=0.7, x = 1, y = 0.25]{A}
\Vertex[x = 2,y = 0.25,style = {color=white}]{D}
\Vertex[x = -2.5,y = 0.25,style = {color=white}]{E}
\Vertex[x = -2.5,y = -0.25,style = {color=white}]{F}
\Vertex[label=$U$, shape=rectangle, size=0.7, x = -1.5, y = 0, style={minimum height=1.2cm}]{U}
\Edge(A)(M1)
\Edge[bend = 40](A)(M3)
\Edge(M1)(E)
\Edge(M2)(F)
\Edge(A)(D)
\node at (-0.83, -0.45) {\tiny $i \neq 0 $};
\end{tikzpicture} +
\begin{tikzpicture}[baseline={([yshift=-0.6ex]current bounding box.center)}]
\Vertex[x = 0, y = 0.25, size = 0.7]{M1}
\Vertex[x = 0, y = -0.25, size = 0]{M2}
\Vertex[x = -0.1, y = -0.1, size = 0]{M3}
\Vertex[x = 0.1, y = -0.1, size = 0]{M4}
\Vertex[label=$M$, shape=rectangle, size=0.7, x = 0, y = 0,style={minimum height=1.2cm},RGB,color={253,192,134}]{M}
\Vertex[label=$A$,size=0.7, x = 1, y = 0.25]{A}
\Vertex[label=$A$,size=0.7, x = -2, y = 0.25]{A3}
\Vertex[label=$A^\dagger$,size=0.7, x = -1, y = 0.25]{A2}
\Vertex[x = 2,y = 0.25,style = {color=white}]{D}
\Vertex[x = -3,y = 0.25,style = {color=white}]{B}
\Vertex[x = -3, y = -0.25 ,style = {color=white}]{C}
\Edge(A)(M1)
\Edge[bend = 40](A)(M3)
\Edge[bend = -40](A2)(M4)
\Edge(A)(D)
\Edge(A2)(M1)
\Edge(A3)(A2)
\Edge(A3)(B)
\Edge[bend = 40](A3)(C)
\end{tikzpicture},
\end{split}
\end{equation}
where $U$ is the reference unitary, as defined in the main text. The first term on the second line is a tensor in the tangent space with
\begin{equation}
x = (1-\ket{0}_p\bra{0}_p) U^\dagger M A \sqrt{\Gamma},\hspace{1cm}\text{or} \hspace{1cm} x = \sum_\mu x_\mu B_\mu = \sum_\mu B_\mu \Tr\left(B_\mu^\dagger M A \Gamma\right),
\end{equation}
showing that part of the effect of introducing the operator $M$ is to create a local perturbation on this site. In the remaining term, the tensor $A$ on this site remains unchanged, but a transformed matrix $A^\dagger M A$ now appears on the right side of the on-site tensor $A$, hence why we call it the pull-through equation. We can apply this equation iteratively to progressively move the matrix $M$ all the way to the right end of the chain, generating local tensor fluctuations along the way. Note that after the first iteration, the matrix $M$ moves to acting on the bond index only, but the more general pull-through equation above can nevertheless be applied. Let us now focus on the recursive map
\begin{equation}
    M^{(i+1)} = A^\dagger (M^{(i)}\otimes I_p)A,
\end{equation}
that describes how the matrix on the bond evolves as it is pulled through the MPS, following the first iteration. The first matrix is $M^{(1)} = A^\dagger M A$, as seen above. Since the physical legs of $A$ and $A^\dagger$ are contracted, we can vectorize this equation to take the form $(M^{(i+1)}| = (M^{(i)}|T(A,A)$, with $T(A,A)$ the transfer matrix of the reference MPS, following the definition in the main text. This makes it clear that for large $i$ we should have $M^{(i)} = \Tr (A^\dagger M A\Gamma) + \mathcal{O}(\lambda^i)$, with $\Gamma$ the right fixed point of the transfer matrix and $\lambda$ the second largest eignevalue of the transfer matrix (the largest eigenvalue being unity for a normalised state). The term that does not decay represents a contribution in the expansion from the reference MPS itself, and can be removed by centering $M$ from the beginning as $M \to M - \Tr (A^\dagger M A\Gamma)$. This shows that the modification to the MPS induced by introducing the operator $M$ can be exactly described as a vector in the first tangent space (+ a component along the reference state if $M$ is not centered). Additionally, this expansion is quasi-local, meaning that terms with derivatives far to the right from the initial insertion site have exponentially small contributions, decaying like $e^{-i/\zeta}$, where $\zeta$ is the correlation length of the reference MPS. Summing all contributions, we see that an operator $h$ acting only on the physical leg at position $i=0$ has a derivative expansion given by
\begin{align}
\label{eq: noderiv}
    h \ket{\Psi} &= \Tr(A^\dagger h A\Gamma) \ket{\Psi} + \sum_{\mu, i\geq 0} x_\mu^{(i)} \ket{\partial_\mu^{(i)} \Psi}, \\
    x^{(i)}_\mu &= \Tr(B_\mu^\dagger h^{(i)} A \Gamma), \\
    h^{(i+1)} &= A^\dagger h^{(i)} A,\hspace{5pt} h^{(0)} = h.
\end{align}
where we have shown that $x_\mu^{(i)}$ decays at a rate given by the MPS inverse correlation length. 

We have so far not accounted for situations where the operator $h$ is applied to states which already contain excitations, i.e. when the local tensor we must pull-through is not $A$, but one of the derivative tensors $B_\mu$. This situation can be handled similarly via a pull-through equation, where the reference tensor $A$ is replaced by the derivative tensor $B_\mu$ introduced in Eq.(4) of the main text. To describe this case it is convenient to define the operator $D_i(M)$, acting on the basis of MPS derivatives by introducing the matrix $M$ on the input legs of the tensor at site $i-1$ and $i$. For convenience we will also assume that $M$ has been centered $\Tr(M\Gamma) = 0$. In diagrammatic form we have

\begin{equation}
D_i(M)\ket{\prod_j \partial_{\nu_j}^{(j)} \Psi} = \begin{tikzpicture}[baseline={([yshift=-0.6ex]current bounding box.center)}]
\Vertex[x = -0.3,y = -0.3,style = {color=white}, size = 0]{M1}
\Vertex[x = -0.7,y = -1,style = {color=white}, size = 0]{M2}
\Vertex[x = 0,y = -0.5,style = {color=white}, size = 0]{M3}
\Vertex[label=$M$, shape=rectangle, size=0.7, x = 0, y = -0.25,style={minimum height=1.2cm},RGB,color={253,192,134}]{M}
\Vertex[label=$B_{\nu_i}$,size=1, x = 1.25, y = 0]{A2}
\Vertex[label=$B_{\nu_{i-1}}$,size=1, x = -1.25, y = 0]{A1}
\Vertex[x = -2.5,y = 0,style = {color=white}]{B1}
\Vertex[x = 2.5,y = 0,style = {color=white}]{B2}
\Vertex[x = -1.25,y = -1,style = {color=white},size = 0]{C1}
\Vertex[x = 1.25,y = -1,style = {color=white}, size = 0]{C2}
\Edge(B1)(B2)
\Edge(C1)(A1)
\Edge[bend = -35](M1)(A2)
\Edge[bend = 45](M2)(M3)
\end{tikzpicture},
\end{equation}
where $\{\nu\}$ is a set of indices, one for each site, specifying which tensor is placed on each site. Here, allow the possibility for $\nu_j$ to be null on specific sites, indicating that there is no derivative on that site, so the local tensor is the reference tensor $A$. Similarly to how we proceeded before, we would like to obtain an expansion of the state above in the basis of derivatives on the reference state. More explicitly, we would like to have
\begin{equation}
\label{eq:D_exp}
    D_i(M)\ket{\prod_j \partial_{\nu_j}^{(j)} \Psi} = \sum_{\{\mu\}} C^{(i)}_{\{\mu\}\{\nu\}}(M) \ket{\prod_j \partial_{\mu_j}^{(j)} \Psi},
\end{equation}
where the sum is over the basis set of all possible configurations of derivatives $\{\mu\}$ placed on the reference tensor. Such an expansion can be obtained by applying the pull-through equation, in the same way as before, but taking into account that the tensor on the right is now $B_{\nu}$ instead of $A$
\begin{equation}
\begin{split}
        \begin{tikzpicture}[baseline={([yshift=-0.6ex]current bounding box.center)}]
\Vertex[x = 0, y = 0.25, size = 0.7]{M1}
\Vertex[x = 0, y = -0.25, size = 0]{M2}
\Vertex[x = -0.1, y = -0.1, size = 0]{M3}
\Vertex[label=$M$, shape=rectangle, size=0.7, x = 0, y = 0,style={minimum height=1.2cm},RGB,color={253,192,134}]{M}
\Vertex[label=$B_{\nu}$,size=0.7, x = 1, y = 0.25]{A}
\Vertex[x = -1,y = 0.25,style = {color=white}]{B}
\Vertex[x = -1,y = -0.25,style = {color=white}]{C}
\Vertex[x = 2,y = 0.25,style = {color=white}]{D}
\Edge(A)(M1)
\Edge[bend = 40](A)(M3)
\Edge(M1)(B)
\Edge(M2)(C)
\Edge(A)(D)
\end{tikzpicture} = 
\begin{tikzpicture}[baseline={([yshift=-0.6ex]current bounding box.center)}]
\Vertex[x = 0, y = 0.25, size = 0.7]{M1}
\Vertex[x = 0, y = -0.25, size = 0]{M2}
\Vertex[x = -0.1, y = -0.1, size = 0]{M3}
\Vertex[label=$M$, shape=rectangle, size=0.7, x = 0, y = 0,style={minimum height=1.2cm},RGB,color={253,192,134}]{M}
\Vertex[label=$B_\nu$,size=0.7, x = 1, y = 0.25]{A}
\Vertex[x = 2,y = 0.25,style = {color=white}]{D}
\Vertex[x = -3.5,y = 0.25,style = {color=white}]{E}
\Vertex[x = -3.5,y = -0.25,style = {color=white}]{F}
\Vertex[label=$U$, shape=rectangle, size=0.7, x = -2.5, y = 0, style={minimum height=1.2cm}]{U}
\Vertex[label=$U^\dagger$, shape=rectangle, size=0.7, x = -1, y = 0, style={minimum height=1.2cm}]{U}
\Edge(A)(M1)
\Edge[bend = 40](A)(M3)
\Edge(M1)(E)
\Edge(M2)(F)
\Edge(A)(D)
\node at (-1, -1.25) {Split to $\ket{0}\bra{0}$ and $1-\ket{0}\bra{0}$};
\draw[->] (-1.75, -1) -- (-1.75, -0.5);
\end{tikzpicture}\\
=\begin{tikzpicture}[baseline={([yshift=-0.6ex]current bounding box.center)}]
\Vertex[x = 0, y = 0.25, size = 0.7]{M1}
\Vertex[x = 0, y = -0.25, size = 0]{M2}
\Vertex[x = -0.1, y = -0.1, size = 0]{M3}
\Vertex[label=$U^\dagger M$, shape=rectangle, size=0.9, x = 0, y = 0,style={minimum height=1.2cm},RGB,color={253,192,134}]{M}
\Vertex[label=$B_\nu$,size=0.7, x = 1, y = 0.25]{A}
\Vertex[x = 2,y = 0.25,style = {color=white}]{D}
\Vertex[x = -2.5,y = 0.25,style = {color=white}]{E}
\Vertex[x = -2.5,y = -0.25,style = {color=white}]{F}
\Vertex[label=$U$, shape=rectangle, size=0.7, x = -1.5, y = 0, style={minimum height=1.2cm}]{U}
\Edge(A)(M1)
\Edge[bend = 40](A)(M3)
\Edge(M1)(E)
\Edge(M2)(F)
\Edge(A)(D)
\node at (-0.83, -0.45) {\tiny $i \neq 0 $};
\end{tikzpicture} +
\begin{tikzpicture}[baseline={([yshift=-0.6ex]current bounding box.center)}]
\Vertex[x = 0, y = 0.25, size = 0.7]{M1}
\Vertex[x = 0, y = -0.25, size = 0]{M2}
\Vertex[x = -0.1, y = -0.1, size = 0]{M3}
\Vertex[x = 0.1, y = -0.1, size = 0]{M4}
\Vertex[label=$M$, shape=rectangle, size=0.7, x = 0, y = 0,style={minimum height=1.2cm},RGB,color={253,192,134}]{M}
\Vertex[label=$B_\nu$,size=0.7, x = 1, y = 0.25]{A}
\Vertex[label=$A$,size=0.7, x = -2, y = 0.25]{A3}
\Vertex[label=$A^\dagger$,size=0.7, x = -1, y = 0.25]{A2}
\Vertex[x = 2,y = 0.25,style = {color=white}]{D}
\Vertex[x = -3,y = 0.25,style = {color=white}]{B}
\Vertex[x = -3, y = -0.25 ,style = {color=white}]{C}
\Edge(A)(M1)
\Edge[bend = 40](A)(M3)
\Edge[bend = -40](A2)(M4)
\Edge(A)(D)
\Edge(A2)(M1)
\Edge(A3)(A2)
\Edge(A3)(B)
\Edge[bend = 40](A3)(C)
\end{tikzpicture},
\end{split}
\end{equation}
such that the first term modifies the derivative tensor from $B_{\nu}$ to $B_{\mu}$ with coefficient
\begin{equation}
    x_\nu = \Tr(B_\mu^\dagger M B_\nu \Gamma),
\end{equation}
and leaves the rest of the chain untouched, while the second term removes the on-site derivative completely, leaving behind the reference tensor $A$, and moves the transformed matrix $M' = A^\dagger M B_\nu$ to the next bond. Though tedious, it is straight-forward to see that applying this method recursively will give us an explicit form for the coefficients $C_{\{\mu\}}^{\{\nu\}}$ as linear functions of the matrix $M$, analogous to Eq.~\eqref{eq: noderiv}.

It is now the moment to point out a difficulty in making sense of the expansion we have performed so far, which is that $D_i(M)$ cannot be interpreted as a linear operator acting in the Hilbert space of physical indices of the MPS. A simple way to see this is to note that our definitions for the local derivative tensors $B_{\mu_i}$ allow for the creation of states of zero norm. It is possible to construct derivative tensors $B_1 = \sum_{\mu} x_{\mu} B_\mu$ and $B_2 = \sum_{\mu} y_{\mu} B_\mu$ whose contraction along a common bond gives identically $0$. Then the entire MPS including this combination should also be identically $0$, and acting with any operator on its physical indices will also give a null result. However, we find that by injecting some matrix $M$ on a contracted bond, it is possible to obtain an MPS which is not null, proving that $D(M)$ is not a well-defined linear operator. The source of this difficulty is that the basis of local derivatives on top of a reference MPS, as described above, is not linearly independent, as we are using $(D^2(d-1)+1)^N$ vectors to span a $d^N$ dimensional Hilbert space.

The solution we propose is to consider a virtual Hilbert space where the basis constructed above by taking derivatives of the MPS is orthonormal. To distinguish them from their associated states in the physical Hilbert space, we will denote states in the virtual space using square bracket notation, such as $|\Psi]$ for the ground state. States in this space will be labeled by sets of local tensor derivative indices $\{\mu\}$, where no more than a single derivative is placed on each site, but some sites can have none. We then define an inner product on this virtual space by imposing orthonormality of the basis states constructed above
\begin{equation}
    [\partial_{\{\mu\}} \Psi | \partial_{\{\nu\}} \Psi] = \delta_{\{\mu\} \{\nu\}}.
\end{equation}

We can now define a pair of bosonic creation and annihilation operators $(a^{\dagger(i)}_\mu,a^{(i)}_\mu)$, obeying the usual commutation relatiosn, for each site $i$ and tensor fluctuation mode $\mu$. They act by adding or removing their respective index to the list $\{\mu\}$ labelling the states, and have $|\Psi]$ as vacuum. While this formally expands the virtual space to include states with more than a single derivative per site, we will see that these become dynamically isolated by the way we construct representations of physical operators in virtual space. It can be checked that this is analogous to the traditional bosonic Holstein-Primakoff transformation, where the operator mapping ensures unphysical states do not become populated.

We can use the virtual space to give a proper definition of the insertion operators $D_i(M)$, using the coefficients in Eq.~\eqref{eq:D_exp}
\begin{equation}
    D_i(M)|\partial_{\{\nu\}} \Psi] = \sum_{\{\mu\}} C^{(i)}_{\{\mu\}\{\nu\}}(M) |\partial_{\{\mu\}} \Psi].
\end{equation}

In practice, we will not work with the coefficients $C^{(i)}_{\{\mu\}\{\nu\}}(M)$ directly, but rather construct an expression in terms of bosonic creation and annihilation operators that produces the same expansion. Careful application of the pull-through equation reveals that this expression should be
\begin{equation}
\begin{split}
    D_i(M) &= \sum_{\mu\nu}a_\mu^{(i)\dagger}a_\nu^{(i)} \Tr\left(\Gamma B_\mu^\dagger M B_\nu\right) + \sum_\mu a_\mu^{(i)\dagger}(1-\sum_\nu a_\nu^{(i)\dagger}a_\nu^{(i)}) \Tr(\Gamma B_\mu^\dagger M A) \\
    &+ \sum_\nu D_{i+1}(A^\dagger M B_\nu)a_\nu^{(i)} + D_{i+1}(A^\dagger M A)(1-\sum_\nu a_\nu^{(i)\dagger}a_\nu^{(i)}).
\end{split}
\end{equation}

There are 4 terms in the expression for the 4 possible terms appearing in the pull-through equation, 2 for when there is already a derivative $\nu$ on site $i$ and 2 for when there is no derivative, which is checked through the $1-\sum_\nu a_\nu^{(i)\dagger}a_\nu^{(i)}$ projector.

If the transfer matrices of the reference state are well-behaved, the series above should remain quasi-local. The condition $\Tr(K\Gamma) = 0$ ensures that performing $K \to A^\dagger K A$ many times leads to exponential decay to 0. It should be noted that the expression above ensures the correct behaviour for all states in the virtual space that have up to a single derivative per site, and it can be easily checked that this subspace is closed. If there are no derivatives, the expression above reduces to Eq.~\eqref{eq: noderiv}. We can now write the Holstein-Primakoff mapping to MPS bosons by expressing the local operator $h$ at site $i$ in terms of $D_{i+1}$ via another pull-through equation. If we assume the expectation value of $h$ is null $\Tr(A^\dagger I\otimes h A\Gamma) = 0$ then

\begin{equation}
\begin{split}
    h &\cong \sum_\mu a_\mu^{(i)\dagger}a_\nu^{(i)} \Tr(\Gamma B_\mu^\dagger I\otimes h B_\nu) + \sum_\mu a_\mu^{(i)\dagger}(1-\sum_\nu a_\nu^{(i)\dagger}a_\nu^{(i)}) \Tr(\Gamma B_\mu^\dagger I\otimes h A) \\
    &+ D_{i+1}(A^\dagger I\otimes h A)(1-\sum_\nu a_\nu^{(i)\dagger}a_\nu^{(i)}) + \sum_\nu D_{i+1}(A^\dagger I\otimes h B_\nu) a_\nu^{(i)},
\end{split}
\end{equation}
where $\cong$ signifies that the same state is obtained by applying $h$ to a tensor directly, or mapping the state to its equivalent in virtual space, applying the bosonic operator on the RHS, and projecting the state back to physical space.

This expansion, alongside the bosonic formulation of the Gram operator described in Sec.~\ref{sec:bosonicmapping} above, is sufficient to reformulate the eigenvalue problem of any physical Hamiltonian $H$ into a generalized eigenvalue problem in the virtual bosonic space. The role of the Gram operator is to recover the physical inner product of states in the tensor derivative basis. This means we should have by definition
\begin{equation}
    \bra{\partial_{\{\mu\}} \Psi}\ket{\partial_{\{\nu\}} \Psi} = [\partial_{\{\mu\}} \Psi |\mathcal{G}| \partial_{\{\nu\}} \Psi].
\end{equation}
Using both this fact and the identification of the physical Hamiltonian $H$ with a bosonic operator $H_B$ constructed as above, we see that the minimization problem
\begin{equation}
    E_{GS} = \min_\phi \frac{\bra{\phi}H\ket{\phi}}{\bra{\phi}\ket{\phi}},
\end{equation}
for the ground state of $H$ will give the same result as the minimization problem
\begin{equation}
    E_{GS} = \min_\phi \frac{[\phi|\mathcal{G}H_B|\phi]}{[\phi|\mathcal{G}|\phi]},
\end{equation}
where $|\phi]$ is a bosonic state. It is known that the solution of this variational problem is given by the solution to the generalized eigenvalue equation
\begin{equation}
    \mathcal{G}H_B|\phi] = E\mathcal{G}|\phi], \hspace{1cm} \mathcal{G}|\phi] \neq 0.
\end{equation}

If an efficient method for obtaining the projector $\mathcal{P} = \mathcal{G}^+\mathcal{G}$ onto the image of $\mathcal{G}$ becomes available, then transforming all bosonic expansion as $h \to \mathcal{P}h\mathcal{P}$ ensures no unphysical states are created by the mapping, in which case the eigenvalue problem takes the simpler form
\begin{equation}
    \mathcal{P}H_B\mathcal{P} |\phi] = E |\phi], \hspace{1cm} \mathcal{P}|\phi] = |\phi],
\end{equation}
and one no longer has to worry about $\mathcal{G}$ not being full rank.

We speculate it may be possible to construct an expression for $\mathcal{P}$ by considering the geometric origin of the Gram operator, but we are unable to do so at the present time.

\end{document}